\documentclass{PoS}
\usepackage{cite}
\usepackage[latin1,utf8]{inputenc}

\newcommand{\bq}{\begin{eqnarray}}
\newcommand{\eq}{\end{eqnarray}}
\newcommand{\eps}{\varepsilon}

\title{Simple differential equations for Feynman integrals associated to elliptic curves}

\ShortTitle{Simple differential equations}

\author{\speaker{Stefan Weinzierl} 
        \\
        Johannes Gutenberg-Universit\"at Mainz\\
        E-mail: \email{weinzierl@uni-mainz.de}}

\abstract{
The $\varepsilon$-form of a system of differential equations for Feynman integrals 
has led to tremendeous progress in our abilities to compute Feynman integrals, as long as they fall
into the class of multiple polylogarithms.
It is therefore of current interest, if these methods extend beyond the case of multiple polylogarithms.
In this talk I discuss Feynman integrals, which are associated to elliptic curves and their differential equations. 
I show for non-trivial examples how the system of differential equations can be brought into an $\varepsilon$-form. 
Single-scale and multi-scale cases are discussed.
}

\FullConference{14th International Symposium on Radiative Corrections (RADCOR2019)\\ 
9-13 September 2019\\
		Palais des Papes, Avignon, France}

\begin{document}


\section{Introduction}

Integration-by-parts identities \cite{Tkachov:1981wb,Chetyrkin:1981qh}
and differential equations \cite{Kotikov:1990kg,Kotikov:1991pm,Remiddi:1997ny,Gehrmann:1999as,Argeri:2007up,MullerStach:2012mp,Henn:2013pwa,Henn:2014qga,Ablinger:2015tua,Adams:2017tga,Bosma:2017hrk}
are standard tools for the computation of Feynman integrals.
In essence, integration-by-parts identities allow us to express a Feynman integral from a large set of Feynman integrals as a linear
combination of Feynman integrals from a smaller set. 
The Feynman integrals in the smaller set are called master integrals
and we may think of the master integrals as a basis of an (abstract) vector space.
We denote the number of master integrals by $N_F = N_{\mathrm{Fibre}}$ and the master integrals by $I = (I_1, ..., I_{N_F})$.
Public available computer programs 
based on the Laporta algorithm \cite{Laporta:2001dd}
like 
\verb|REDUZE| \cite{vonManteuffel:2012np},
\verb|FIRE| \cite{Smirnov:2014hma} or
\verb|KIRA| \cite{Maierhoefer:2017hyi}
can be used to perform the reduction to the master integrals.

For the master integrals one derives (again by using integration-by-parts identities)
differential equations in the external invariants or internal masses.
We denote the number of kinematic variables by $N_B = N_{\mathrm{Base}}$ and the kinematic variables by $x=(x_1, ..., x_{N_B})$.
The system of differential equations for the master integrals can be written as
\bq
\label{eq1}
 d I + A I & = & 0,
\eq
where $A(\eps,x)$ is a matrix-valued one-form
\bq
 A & = & 
 \sum\limits_{i=1}^{N_B} A_i dx_i.
\eq
The $A_i(\eps,x)$'s are matrices of size $N_F \times N_F$, whose entries are rational functions in the dimensional regularisation parameter $\eps$
and the kinematic variables $x$.
The matrix-valued one-form $A$ satisfies the integrability condition
\bq
 dA + A \wedge A & = & 0.
\eq
Geometrically we have a vector bundle with a fibre of dimension $N_F$ spanned by $I_1, \dots I_{N_F}$ 
and a base of dimension $N_B$ with local coordinates $x_1, \dots, x_{N_B}$.
The matrix-valued one-form $A$ defines a flat connection.

Up to this point everything is general and applies to any Feynman integral.
In particular, computing a Feynman integral is reduced to the problem of solving a system of differential equations as in eq.~(\ref{eq1}).
The solution of a system of differential equations requires in addition boundary values.
The boundary values correspond to simpler Feynman integrals, where some kinematic variables have special values or vanish.
Therefore at this stage the boundary values can be considered to be known (otherwise one would first set up a system of differential
equations for the boundary values).

The system of differential equations is particular simple \cite{Henn:2013pwa}, if $A$ is of the form
\bq
\label{eq2}
 A & = &
 \eps \; \; \sum\limits_{j=1}^{N_L} \; C_j \; \omega_j,
\eq
where
\begin{description}
\item{-} the only dependence on the dimensional regularisation parameter $\eps$ is given by the explicit prefactor,
\item{-} the $C_j$'s are $N_F \times N_F$-matrices, whose entries are (rational or integer) numbers,
\item{-} the differential one-forms $\omega_j$ have only simple poles (and depend only on $x$).
\end{description}
We denote by $N_L = N_{\mathrm{Letters}}$ the number of letters, i.e. the number of ${\mathbb Q}$-linear independent differential
one-forms $\omega_j$. The set of letters is denoted by $ \omega=(\omega_1, ..., \omega_{N_L})$.

A system of differential equations in the form of eq.~(\ref{eq2}) is easily solved order-by-order
in the dimensional regularisation parameter $\eps$ in terms of
iterated integrals.
For $\omega_1$, ..., $\omega_j$ differential 1-forms on a manifold $B$
and $\gamma : [0,1] \rightarrow B$ a path,
let us write for the pull-back of $\omega_i$ to the interval $[0,1]$
\bq
 f_i\left(\lambda\right) d\lambda & = & \gamma^\ast \omega_i.
\eq
The iterated integral is defined by
\cite{Chen}
\bq
 I_{\gamma}\left(\omega_1,...,\omega_j;\lambda\right)
 & = &
 \int\limits_0^{\lambda} d\lambda_1 f_1\left(\lambda_1\right)
 \int\limits_0^{\lambda_1} d\lambda_2 f_2\left(\lambda_2\right)
 ...
 \int\limits_0^{\lambda_{j-1}} d\lambda_j f_j\left(\lambda_j\right).
\eq
We see that the computation of any Feynman integrals is reduced to finding a transformation (if it exists) 
of the system of differential equations to the simple form
of eq.~(\ref{eq2}).

A special case of iterated integrals are multiple polylogarithms.
Assume that all $\omega_j$'s are of the form
\bq
 \omega_j & = & d \ln p_j\left(x\right),
\eq
where the $p_j$'s are polynomials in the variables $x$, then (after factorisation of univariate polynomials)
\bq
 f_i & = & \frac{d\lambda}{\lambda - z_i}
\eq
and all iterated integrals are multiple polylogarithms:
\bq
G(z_1,...,z_j;\lambda) & = & \int\limits_0^\lambda \frac{d\lambda_1}{\lambda_1-z_1}
 \int\limits_0^{\lambda_1} \frac{d\lambda_2}{\lambda_2-z_2} ...
 \int\limits_0^{\lambda_{j-1}} \frac{d\lambda_j}{\lambda_j-z_j}.
\eq
Let us now discuss the possibilities to transform a generic system of differential equations as in eq.~(\ref{eq1})
into the simple form of eq.~(\ref{eq2}).
On the one hand we may change 
change the basis of the master integrals
\bq
 I' & = & U I,
\eq
where $U(\eps,x)$ is a $N_F \times N_F$-matrix.
The new connection matrix is
\bq
 A' & = & U A U^{-1} + U d U^{-1}.
\eq
On the other hand, we may perform a coordinate transformation on the base manifold:
\bq
 x_i' & = & f_i\left(x\right), \;\;\;\;\;\;\;\;\; 1 \le i \le N_B.
\eq
The connection transforms as 
\bq
 A \; = \; \sum\limits_{i=1}^{N_B} A_i dx_i
 & \;\;\;\;\;\; \Rightarrow \;\;\;\;\;\; &
 A' \; = \; \sum\limits_{i,j=1}^{N_B} A_i \; \frac{\partial x_i}{\partial x_j'} \; dx_j'.
\eq
Let us stress 
\begin{figure}
\begin{center}
\includegraphics[scale=1.0]{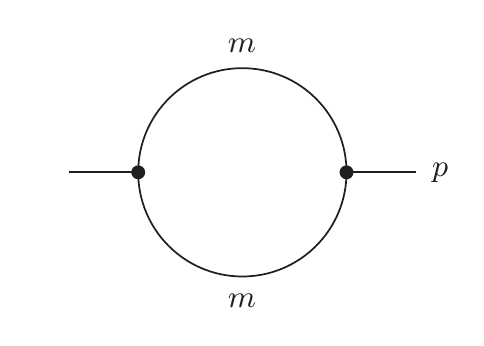}
\end{center}
\caption{
A one-loop two-point integral.
}
\label{fig_bubble}
\end{figure}
that a coordinate transformation on the base manifold is already required in rather simple cases.
Consider the one-loop two-point function with an equal internal mass shown in fig.~\ref{fig_bubble}.
Transforming the system of differential equations into a form, where $\eps$ appears only as a prefactor will
inevitably introduce the square root
\bq
 \frac{dx}{\sqrt{-x\left(4-x\right)}},
\eq
where $x=p^2/m^2$.
Here, a change of variables in the base manifold
\bq
 x & = & - \frac{\left(1-x'\right)^2}{x'}
\eq
will rationalise the square root and transform
\bq
 \frac{dx}{\sqrt{-x\left(4-x\right)}}
 & = &
 \frac{dx'}{x'}.
\eq
Can the required transformations be found systematically?
In the case of Feynman integrals evaluating to multiple polylogarithms
there are systematic algorithms to find a transformation of the basis of master integrals $I' = U I$
provided that $U$ is rational in the 
kinematic variables \cite{Henn:2013pwa,Gehrmann:2014bfa,Argeri:2014qva,Lee:2014ioa,Prausa:2017ltv,Gituliar:2017vzm,Meyer:2016slj,Meyer:2017joq,Lee:2017oca}.
In the case of coordinate transformations on the base manifold $x_i' = f_i\left(x\right)$
there are now systematic algorithms to rationalise square roots \cite{Becchetti:2017abb,Besier:2018jen,Besier:2019aaa}.

With the help of these algorithms the simple form of the differential equations as in eq.~(\ref{eq2})
can be reached for many Feynman integrals evaluating to multiple polylogarithms.
Please note that these algorithms still have some limitations.
Not all Feynman integrals, which can be expressed in terms of multiple polylogarithms,
can be treated with the algorithms mentioned above.
An example where further technical improvements are desirable is given by 
the two-loop electroweak-QCD corrections to the Drell-Yan process \cite{Heller:2019gkq}.

However, not all Feynman integrals can be expressed in terms  of multiple polylogarithms.
The next more complicated case are Feynman integrals associated to elliptic curves.
These are a topic of current research 
interest \cite{Laporta:2004rb,MullerStach:2011ru,Adams:2013nia,Bloch:2013tra,Remiddi:2013joa,Adams:2014vja,Adams:2015gva,Adams:2015ydq,Bloch:2016izu,Adams:2017ejb,Bogner:2017vim,Adams:2018yfj,Honemann:2018mrb,Bloch:2014qca,Sogaard:2014jla,Tancredi:2015pta,Primo:2016ebd,Remiddi:2016gno,Adams:2016xah,Bonciani:2016qxi,vonManteuffel:2017hms,Adams:2017tga,Ablinger:2017bjx,Primo:2017ipr,Passarino:2017EPJC,Remiddi:2017har,Bourjaily:2017bsb,Hidding:2017jkk,Broedel:2017kkb,Broedel:2017siw,Broedel:2018iwv,Lee:2017qql,Lee:2018ojn,Adams:2018bsn,Adams:2018kez,Broedel:2018qkq,Bourjaily:2018yfy,Bourjaily:2018aeq,Besier:2018jen,Mastrolia:2018uzb,Ablinger:2018zwz,Frellesvig:2019kgj,Broedel:2019hyg,Blumlein:2019svg,2019arXiv190611857B,Bogner:2019lfa,Frellesvig:2019uqt,Kniehl:2019vwr,Broedel:2019kmn,Duhr:2019rrs}
and the focus of this talk.
We may ask if the simple form for the system of differential equations as in eq.~(\ref{eq2}) can also be obtained
in these cases.
The results obtained so far look promising. 
In the following we will discuss how the form of eq.~(\ref{eq2}) is obtained for the equal mass sunrise integral
and the unequal mass sunrise integral.
In both systems only one elliptic curve occurs.
The former integral depends on one kinematic variable ($N_B=1$), while the latter depends on three kinematic variables ($N_B=3$).
We expect the methods and techniques used in these examples to carry over to the wider class of multi-scale Feynman integrals
associated with a single elliptic curve.
An example where exactly the same techniques can be applied would be the kite integral \cite{Remiddi:2016gno,Adams:2016xah}. This integral is relevant to
the two-loop electron self-energy in QED \cite{Honemann:2018mrb}.

\section{One elliptic curve, one variable}

Let us start with the single scale case. The standard example is the equal mass sunrise integral.
We have $3$ master integrals and one kinematic variable, i.e. $N_F=3$, $N_B=1$. 
As kinematic variable we use $x = p^2/m^2$.
The first question which we should address is how to obtain the elliptic curve associated to this integral.
For the sunrise integral there are two possibilities, we may either obtain an elliptic curve from the Feynman graph polynomial
\bq
 - x_1 x_2 x_3 x + \left( x_1 + x_2 + x_3 \right) \left( x_1 x_2 + x_2 x_3 + x_3 x_1 \right) 
 & = & 0,
\eq
where $x_1,x_2,x_3$ are Feynman parameters and $x$ the kinematic variable or from the maximal cut:
\bq
\label{maxcut}
 v^2 
 -
 \left(u - x \right) 
 \left(u - x + 4 \right) 
 \left(u^2 + 2 u + 1 - 4 x \right)
 & = & 0.
\eq
Please note that these two elliptic curves are not isomorphic, but only isogenic.
The Weierstrass normal form of an elliptic curve reads
\bq
\label{WNF}
 v^2 & = & 4 \left(u-e_1\right) \left(u-e_2\right) \left(u-e_3\right),
\eq
where we already factorised the cubic polynomial in $u$ on the right-hand side.
The periods of the elliptic curve in eq.~(\ref{WNF}) are given by
\bq
 \psi_1 = 2 \int\limits_{e_1}^{e_2} \frac{du}{v} = \frac{2}{\sqrt{e_3-e_1}} K\left(\sqrt{ \frac{e_2-e_1}{e_3-e_1}}\right),
 & &
 \psi_2 = 2 \int\limits^{e_2}_{e_3} \frac{du}{v} = \frac{2i}{\sqrt{e_3-e_1}} K\left(\sqrt{ \frac{e_3-e_2}{e_3-e_1}}\right),
 \;\;\;
\eq
where $K(k)$ denotes the complete elliptic integral of the first kind.
The periods $\psi_1$, $\psi_2$ of the elliptic curve are solutions of the homogeneous system of eq.~(\ref{eq1}) \cite{Adams:2013nia}.
This holds independently if we start from the Feynman graph polynomial or from the maximal cut.
The periods associated to eq.~(\ref{maxcut}) equal the maximal cuts.
In general, the maximal cuts are solutions of the homogeneous system of eq.~(\ref{eq1}) \cite{Primo:2016ebd}.

In the mathematical literature the shape of the elliptic curve is often described by  the variable $\tau$ (or $q$), defined by
\bq
 \tau \; = \; \frac{\psi_{2}}{\psi_{1}},
 & &
 q \; = \; e^{2 i \pi \tau}.
\eq
The periods $\psi_1$ and $\psi_2$ generate a lattice. 
Any other basis of the lattice is as good as $(\psi_2,\psi_1)$.
One often normalises one basis vector to one, e.g. $(\psi_2,\psi_1) \rightarrow (\tau, 1)$ with $\tau=\psi_2/\psi_1$.
\begin{figure}
\begin{center}
\includegraphics[scale=1.0]{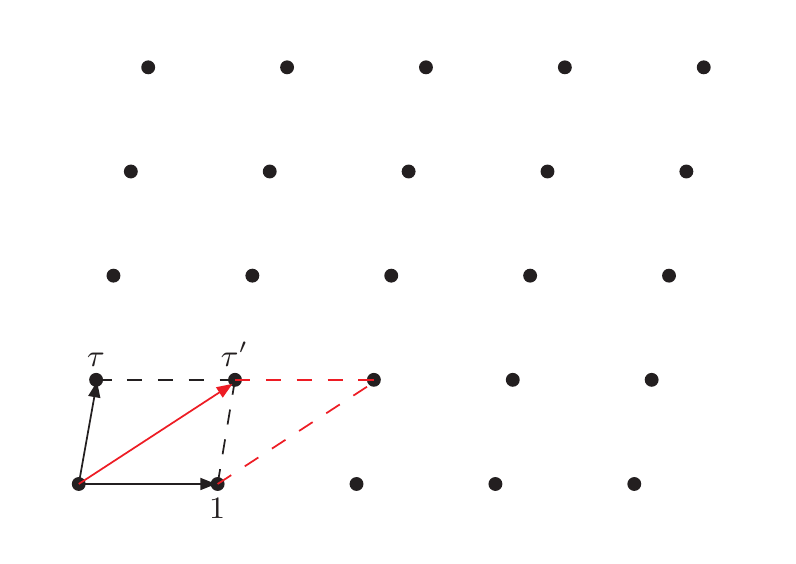}
\end{center}
\caption{
The pair of vectors $1$ and $\tau$ and the pair of vectors $1$ and $\tau'$ generate the same lattice.
}
\label{fig_lattice}
\end{figure}
Let us now consider a change of basis for the basis vectors of the lattice:
\bq
\label{modular_trafo}
 \left( \begin{array}{c}
 \psi_2' \\
 \psi_1' \\
 \end{array} \right)
 & = &
 \left( \begin{array}{cc}
 a & b \\
 c & d \\
 \end{array} \right)
 \left( \begin{array}{c}
 \psi_2 \\
 \psi_1 \\
 \end{array} \right),
 \;\;\;\;\;\;\;\;\;
 \gamma \; = \;
 \left( \begin{array}{cc}
 a & b \\
 c & d \\
 \end{array} \right).
\eq
The transformation should be invertible, therefore $\gamma \in \mathrm{SL}\left(2,{\mathbb Z}\right)$.
In terms of $\tau$ and $\tau'$ we have
\bq
 \tau' & = & \gamma\left(\tau\right) \; = \; \frac{a \tau +b}{c \tau +d}.
\eq
Transformations as in eq.~(\ref{modular_trafo}) are called modular transformations.
A function $f(\tau)$ is called a modular form if
\bq
 f\left(\gamma\left(\tau\right)\right)
 & = &
 \left(c \tau + d\right)^k f\left(\tau\right),
\eq
and $f(\tau)$ is holomorphic in the complex upper half-plane ${\mathbb H}$ and at the cusp $\tau=i\infty$.
The number $k$ is called the modular weight of $f(\tau)$.

In order to bring the system of differential equations for the equal mass sunrise integral into the simple form of eq.~(\ref{eq2})
we perform a change of the basis of the master integrals from a pre-canonical basis $(S_{110},S_{111},S_{211})$ to
\bq
 I_1 = 4 \eps^2 S_{110},
 \;\;\;\;\;
 I_2 = \eps^2 \frac{\pi}{\psi_1} S_{111},
 \;\;\;\;\;
 I_3
 = 
 \frac{1}{\eps} \frac{1}{2\pi i} \frac{d}{d\tau} I_2
 +
 \frac{1}{24} \left(3x^2-10x-9\right) \frac{\psi_1^2}{\pi^2} I_2.
\eq
This transformation is not rational or algebraic in $x$, as can be seen from the prefactor $1/\psi_1$ in the definition of $I_2$.
The period $\psi_1$ is a transcendental function of $x$.
In addition we change the kinematic variable from $x$ to $\tau$ (or $q$) \cite{Bloch:2013tra}.
Again, this is a non-algebraic change of variables.
One obtains
\bq
 d I
 \; = \;
 \eps \; A \; I,
 \;\;\;
 & &
 \;\;\;
 A \; = \; A_\tau d\tau,
\eq
The $\eps$-independent $3 \times 3$-matrix
$A_\tau$ is given by
\bq
 A_\tau & = & 
 \left( \begin{array}{rrr}
 0 & 0 & 0 \\
 0 & -f_2(\tau) & 1 \\
 \frac{1}{4} f_3(\tau) & f_4(\tau) & -f_2(\tau) \\
 \end{array} \right),
\eq
where $f_2$, $f_3$ and $f_4$ are modular forms of $\Gamma_1(6)$ of modular weight $2$, $3$ and $4$, respectively.
This allows us to express $I_1$, $I_2$ and $I_3$ as iterated integrals of modular forms to all orders in $\eps$ \cite{Adams:2017ejb,Adams:2018yfj}.
A modular form $f_k(\tau)$ is by definition holomorphic at the cusp and has a $q$-expansion
\bq
 f_k(\tau) & = & a_0 + a_1 q + a_2 q^2 + ...,
 \;\;\;\;\;\;\;\;\;\;\;\;
 q=\exp(2\pi i \tau).
\eq
The transformation $q=\exp(2\pi i \tau)$ transforms the point $\tau=i\infty$ to $q=0$ and we have
\bq
 2 \pi i f_k(\tau) d\tau & = & \frac{dq}{q} \left( a_0 + a_1 q + a_2 q^2 + ... \right).
\eq
Thus a modular form non-vanishing at the cusp $\tau=i\infty$ has a simple pole at $q=0$.


\section{One elliptic curve, several variables}

Let us now consider the multi-scale case.
Our standard example is the unequal mass sunrise integral.
We have $7$ master integrals and $3$ kinematic variables, i.e. $N_F=7$, $N_B=3$.
As kinematic variables we use
$x = p^2/m_3^2$,
$y_1 = m_1^2/m_3^2$,
$y_2 = m_2^2/m_3^2$.
The system of differential equations can again be transformed into the simple form of eq.~(\ref{eq2}) by a redefinition of the master integrals 
and a change of coordinates.
The explicit formula for fibre transformation is a little bit lengthy and can be found in the literature \cite{Bogner:2019lfa}.
Let us discuss here the coordinate transformation. It has a nice geometric interpretation.
We introduce the moduli space ${\mathcal M}_{g,n}$ as the space of isomorphism classes of smooth (complex, algebraic) 
curves of genus $g$ with $n$ marked points.
\begin{figure}
\begin{center}
\includegraphics[scale=1.0]{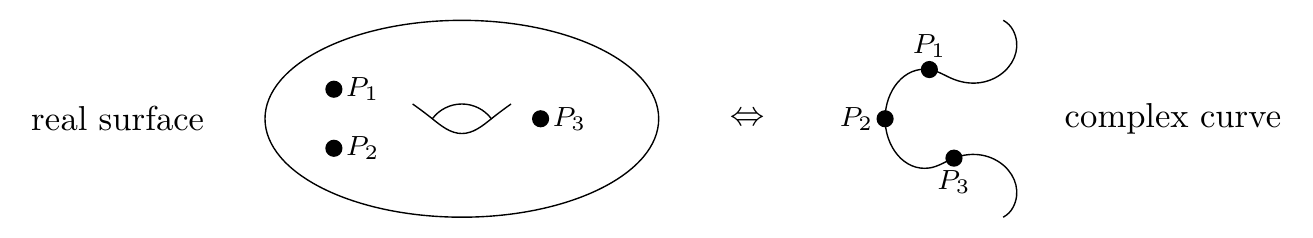}
\end{center}
\caption{
A complex curve can be viewed as a real surface.
}
\label{fig_torus}
\end{figure}
Please note that a complex curve can also be viewed as a real (Riemann) surface (fig.~\ref{fig_torus}).
The dimension of ${\mathcal M}_{g,n}$ is
\bq
 \dim {\mathcal M}_{g,n} & = & 3 g + n - 3.
\eq
Let us now introduce coordinates on ${\mathcal M}_{g,n}$.
We are interested in the cases $g=0,1$.
In genus $0$ we have $\dim {\mathcal M}_{0,n} = n - 3$. The Riemann sphere has a unique shape. 
The isomorphisms are the M\"obius transformations and we may use a M\"obius transformation
to fix three marked points at specific values, say $z_{n-2}=1$, $z_{n-1}=\infty$, $z_n=0$.
Thus we may take $(z_1,...,z_{n-3})$ as coordinates on ${\mathcal M}_{0,n}$.

In genus $1$ we have $\dim {\mathcal M}_{1,n} = n$. We need one coordinate to describe the shape of the torus.
As isomorphisms we only have translations, which can be used to fix $z_n=0$.
Thus we may take $(\tau,z_1,...,z_{n-1})$ as coordinates on ${\mathcal M}_{1,n}$.

For the unequal mass sunrise integral we change coordinates from $(x,y_1,y_2)$ to coordinates $(\tau,z_1,z_2)$ of ${\mathcal M}_{1,3}$.
(In this language the change of coordinates for the equal mass sunrise integral is from the coordinate $x$ to the coordinate $\tau$
of ${\mathcal M}_{1,1}$.)
This raises the question how to find $z_1$ and $z_2$. The coordinate $\tau$ is defined as before as the ratio $\psi_2/\psi_1$.
In the Feynman parameter representation there are two geometric objects of interest:
the domain of integration $\sigma$ and the zero set $X$ of the second graph polynomial.
\begin{figure}
\begin{center}
\includegraphics[scale=1.0]{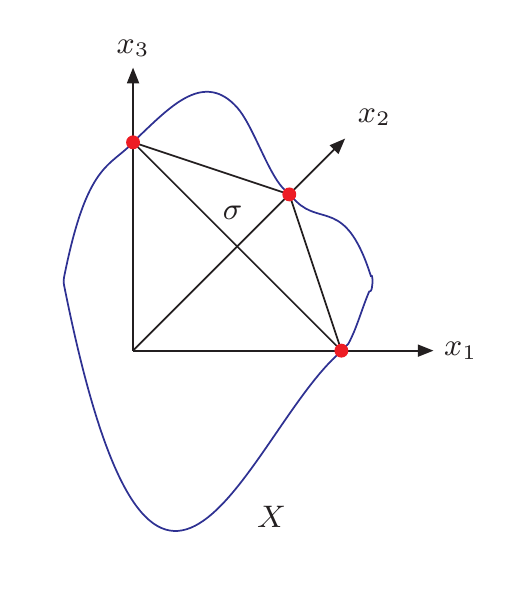}
\includegraphics[scale=1.0]{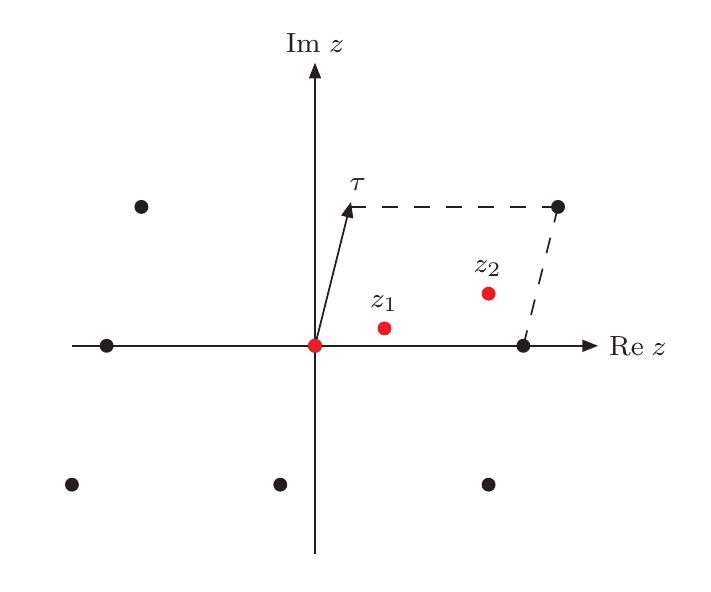}
\end{center}
\caption{
$X$ and $\sigma$ intersect at three points, the images of these three points in ${\mathbb C}/\Lambda$ are $0,z_1,z_2$.
}
\label{fig_intersection}
\end{figure}
$X$ and $\sigma$ intersect at three points, as shown in fig.~\ref{fig_intersection}.
The images of these three points in ${\mathbb C}/\Lambda$ are $0,z_1,z_2$, where we used a translation transformation to fix one point at $0$.
After a redefinition of the basis of master integrals and a change of coordinates from
$(x,y_1,y_2)$ to $(\tau,z_1,z_2)$ one obtains \cite{Bogner:2019lfa} the simple form of eq.~(\ref{eq2}) 
\bq
 A & = &
 \eps \; \; \sum\limits_{j=1}^{N_L} \; C_j \; \omega_j,
 \;\;\;\;\;\;\;\;\;\;\;\; 
 \mbox{with $\omega_j$ only simple poles,} 
\eq
where $\omega_j$ is either 
\bq
 \left(2\pi\right)^{2-k}
 f_k\left(\tau\right) \frac{d\tau}{2\pi i},
\eq
where $f_k(\tau)$ is a modular form, or of the form
\bq
\omega_k\left(z_i,\tau\right)
 & = &
 \left(2\pi\right)^{2-k}
 \left[
  g^{(k-1)}\left(z_i,\tau\right) dz_i + \left(k-1\right) g^{(k)}\left(z_i,\tau\right) \frac{d\tau}{2\pi i}
 \right],
\eq
where $g^{(k)}(z,\tau)$ are functions appearing in the expansion of the Kronecker function
\bq
F\left(z,\alpha,\tau\right)
 & = &
 \pi
 \vartheta_1'\left(0,q\right) \frac{\vartheta_1\left( \pi\left(z+\alpha\right), q \right)}{\vartheta_1\left( \pi z, q \right)\vartheta_1\left( \pi \alpha, q \right)}
 \; = \;
 \frac{1}{\alpha} \sum\limits_{k=0}^\infty g^{(k)}\left(z,\tau\right) \alpha^k,
 \;\;\;\;\;\;\;\;\;
 q = e^{i \pi \tau}.
\eq
$\vartheta_1$ denotes the first Jacobi $\vartheta$-function.
The properties of $g^{(k)}(z,\tau)$ are
\cite{Brown:2011,Broedel:2017kkb}:
$g^{(k)}$ has only simple poles as a function of $z$,
$g^{(k)}$ is quasi-periodic as a function of $z$, i.e. periodic by $1$ and quasi-periodic by $\tau$,
$g^{(k)}$ is almost modular, the nice modular transformation properties
are only spoiled by the divergent Eisenstein series $E_1(z,\tau)$.

Finally let us remark that the equal mass case corresponds to the situation, where $z_1$ and $z_2$ attain the fixed
value $z_1=z_2=1/3.$


\section{Outlook}

The computation of Feynman integrals is trivial, as soon as the system of differential equations is transformed to
\bq
 A & = &
 \eps \; \; \sum\limits_{j=1}^{N_L} \; C_j \; \omega_j,
 \;\;\;\;\;\;\;\;\;\;\;\; 
 \mbox{with $\omega_j$ only simple poles.} 
\eq
This form can be reached for 
many Feynman integrals evaluating to multiple polylogarithms and
a few non-trivial elliptic examples.
It is an open question if any Feynman integral can be obtained from a system of differential equations of this form.
A constructive proof would gives us an algorithm to compute any Feynman integral.

\bibliography{/home/stefanw/notes/biblio}
\bibliographystyle{/home/stefanw/latex-style/h-physrev5}

\end{document}